\documentclass[a4paper,11pt]{article}
\usepackage{jinstpub} 
\usepackage{lineno}

\title{\boldmath APEX: Optimized vertical drift PDS for DUNE FD3}

\collaboration[c]{on behalf of DUNE collaboration}

\author{F. Marinho}
\affiliation{Departamento de Física, Instituto Tecnológico de Aeronáutica, São José dos Campos, 12228-900 SP, Brazil}

\emailAdd{franciole@ita.br}

\abstract{The Deep Underground Neutrino Experiment currently under construction in the US will be a long-baseline neutrino oscillation experiment dedicated to determining the neutrino mass ordering and to measure the CP violation phase in the lepton sector. DUNE will also perform studies of non-beam physics such as atmospheric neutrinos, bursts from supernovae and nucleon decays. For these kind of interactions, the photon detection systems will play a major role in triggering and also provide calorimetric measurements. For the second phase of DUNE, two additional detector modules will be added in the far detector complex in the Sanford Underground Research Facility. We present the Aluminum Profiles with Embedded X-ARAPUCA (APEX) concept as an advanced proposal for the photon detector system of the third DUNE far detector module. This system aims to have an optical coverage of approximately 60\% made viable by the technology advancement achieved by the DUNE collaboration on the use of non-conductive optical fibers for power and signal readout of the photon detector units. Such large coverage will provide enhanced light collection capabilities at MeV-scale energy deposit level per interaction and optimal energy reconstruction resolution up to the GeV scale. The attained electrical isolation of the detector units with low noise levels allows for a complete instrumentation of the field cage walls with satisfactory segmentation as the readout scheme envisages a much larger than typical number of channels to be adopted. We discuss the main features of the system, first estimates on its expected performances, potential for physics measurements and prototyping plans for R\&D.}


\arxivnumber{} 

\begin{document}
\maketitle
\flushbottom

\section{Introduction}


The Deep Underground Neutrino Experiment is a next-generation neutrino detector dedicated to the study of the oscillation phenomena using a powerful neutrino beam, atmospheric neutrinos, neutrinos from astrophysical sources as well as being capable of performing physics beyond the standard matter searches~\cite{dunevol1}. DUNE has a two-phase implementation plan. Phase I will be composed of two 17 kton far detectors (FD) liquid argon time projection chambers (LArTPCs), a near detector (ND) with a LArTPC module (ND-LAr), a muon spectrometer (TMS), with the addition of an on-axis solenoid magnet plus electromagnetic calorimeter (SAND), and a beam line capable of providing (anti-)neutrinos with a 1.2 MW power. Phase II aims at expanding DUNE capabilities by limiting systematics with a more capable ND and greatly increasing statistics with the construction of another two FDs in addition to boosting beam power to 2.1 MW~\cite{phase2}.

The LArTPC technology provides excellent imaging capabilities and good energy resolution, making DUNE able to fulfill its broad physics program on its own. Its powerful sensitivity to electron neutrinos also complements the measurements to be obtained by other large neutrino detectors. In particular, in phase I DUNE will be able to determine at a 5$\sigma$ level the neutrino mass ordering and study neutrino oscillation parameters \cite{phase2}. Phase II will allow the determination of the $\theta_{23}$ octant and $\delta_{CP}$ value, if not nearly maximal, as well as the investigation of new phenomena.

In phase I, there will be one module with anode plane assemblies (APAs) composed of thin copper wires for charge collection displaced vertically so that the ionization electrons will drift horizontally, therefore receiving the name of the Horizontal Drift (HD) module. The scintillation light collecting system will have bar-shaped X-Arapucas placed in between the APAs~\cite{dunevol4}. The other module will have its anodes composed of perforated PCB layers with conductive strips that will be disposed vertically in the cryostat so that the electric field will be applied vertically. This configuration is called the Vertical Drift (VD) module and its advantageous features are described in reference ~\cite{Vdtdr}. Its Photon Detection System (PDS) will have rectangular X-Arapucas positioned along the cryostat walls (outside the active volume) and also on the cathode, in a high voltage environment. A substantial R\&D was developed by the collaboration in order to have the photon detectors working using signal-over-fiber and power-over-fiber in a cryogenic environment \cite{Arroyave_2024}.

The FDs in phase II will be built profiting from the experience the collaboration have acquired in designing and constructing phase I modules. In particular, the current baseline design of the third module is envisioned to also employ PCB layers with conductive strips for anodes in a vertical drift configuration. Nevertheless, a more capable PDS is being planned for it since an improved scintillation light collection combined with lower background rates carry the potential benefit of lowering the energy threshold for neutrino detection (currently at 5 MeV in phase I) and improving energy resolution. This is of particular importance allowing for better solar neutrino measurements, enhancement of the precision over $\Delta m^2_{12}$ measurement and the investigation of solar metallicity to a finer level \cite{phys_prosp}. In addition, it should improve the determination of the second oscillation peak maximum at $\sim$0.8 GeV \cite{PhysRevD.111.032007} (which presents a lower interaction count due to the dependence of the cross section on neutrino energy and DUNE beam energy spread in comparison with the first oscillation peak around 2.5 GeV). The Aluminum Profiles with Embedded X-ARAPUCA (APEX) concept aligns well with this.


In APEX, the photon detectors are integrated into the field cage allowing for a large coverage of the active volume. Since the field cage covers the four vertical sides of the LArTPC active volume between the top anode and bottom anode planes, it offers the largest available surface for extended
optical coverage. Building on the expertise developed for the cathode PD modules in VD, this concept presents a simplified, lightweight, and low(er)-cost photodetectors, combined with bulk materials of low-radioactive content.



\section{APEX: Description \& Performance}

Established on the idea of embedded X-Arapuca photon detectors on aluminum profiles of its field cage (FC), APEX is a PDS reference concept proposed for the third module of the DUNE experiment \cite{phase2}. This design was naturally further developed based on the achievements reached with the VD technology \cite{Vdtdr} which will be implemented on the first phase of the DUNE also having two drift regions above and below a horizontal cathode at mid height. This configuration makes APEX agnostic with respect to any charge collection system that might be adopted in this detector module. As done in DUNE's second module, the doping of LAr with Xe at the order of 10ppm is also foreseen in APEX due to the large volume that needs to be covered by the photon-detectors.

\begin{figure}[htbp]
\centering
\includegraphics[width=.55\textwidth]{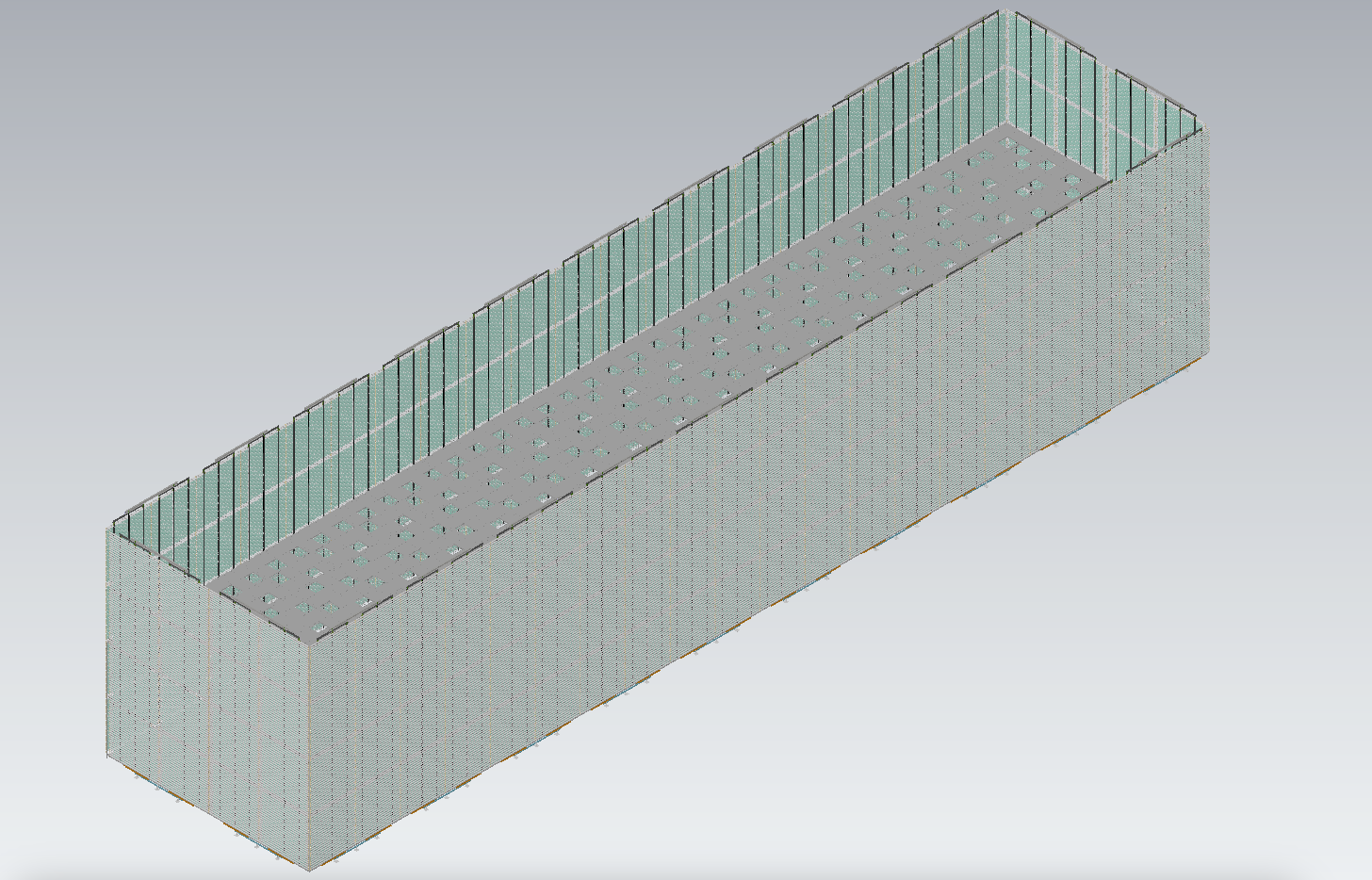}
\includegraphics[width=.275\textwidth]{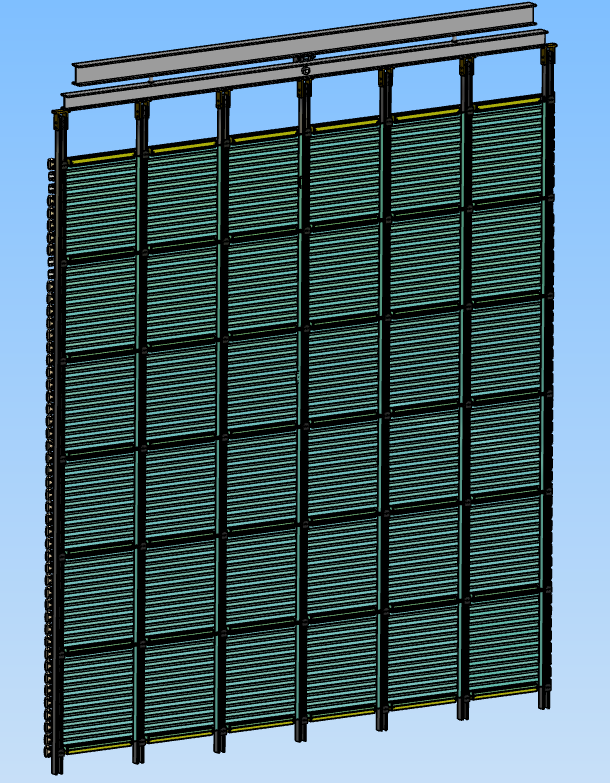}
\caption{A view of the complete APEX PDS concept (left) and one $\rm 6\times6$ full detecting panel (right).\label{fig:apex}}
\end{figure}

In comparison to VD, APEX brings many improvements such that in the second phase of DUNE it should strongly impact the overall experiment performance both in the timing and trigger of non-beam events as well as additional calorimetric measurements of neutrino interactions with primary particle energies ranging from $\rm \sim 5 MeV$ up to tens of GeV \cite{PhysRevD.111.032007, phys_prosp}. These are expected as this system envisages a large optical coverage with $\sim 7000$ PD units with an area of $\rm 50 \times 50~ cm^2$/unit all positioned on the four vertical sides of the FC, being 2 shorter sides ($\rm 12 \times 14 ~m^2$) and 2 longer sides ($\rm 60 \times 14 ~m^2$). This arrangement enhances light yield ($LY$) and uniformity of the PDS response across full volume. Figure \ref{fig:apex} shows a complete view of the APEX system on left panel. The $z$ axis is in the beam orientation, drift is vertical and cathode is horizontally oriented. A panel composed of $\rm 6\times6$ PD units mounted in a standard VD FC structural module is shown on right panel.

Another important development inherited from the VD is the successful employment of power and signal over fiber technology (PoF/SoF) for safely operating readout electronics on HV surfaces of the FC with data routed via non-conductive optical fibers. However, new features such as channels multiplexing through wavelength band division implemented in cold should allow for affordable data acquisition from all PD units at appropriate rates.

In order to evaluate performance of the APEX system a standalone Monte Carlo simulation was implemented using Geant4 \cite{2003Geant4, 2006Geant4, 2016Geant4} allowing mapping of the overall detector response $LY$ and estimates of position and energy resolution at the MeV scale. The complete cryostat volume was considered including features of the cathode, FC profiles, mechanical structures and all PDs. The low energy events in this simulation were generated as simple point-like sources within the sensitive volume as track lengths are very short. Table \ref{tab:Params} lists the adopted optical characteristics for the materials in the cryostat. The $R$ reflectances for each component are included in the simulation. Scintillation yield was $\rm \sim25k$ photons per deposited MeV and the emitted photons wavelengths were 128 nm (Ar) and 176 nm (Xe). LAr refractive index dependency with light wavelength is also implemented. Another relevant aspect that is taken into consideration is the fact that the X-Arapucas PDs for APEX contain a thin layer of p-terphenyl (pTP) acting as a first stage wavelength shifter in its acceptance window. Because of the large PDS optical coverage the amount of the pTP shifted light that is emitted backwards into the cryostat volume has to be accounted for as there is a large probability of being detected by another PD unit.

\begin{table}[h]
\centering
\caption{Simulation optical properties.}
\label{tab:Params}
\begin{tabular}{l|c|c|c|c}
\hline
 Parameter & LAr scintillation & Xe doping & Rayleigh scattering & Absorption length \\ 
  & yield (mip) &  & & \\ \hline
 Value     & 25k ph/MeV & 10 ppm & {\begin{tabular}[c]{@{}c@{}} $\rm \lambda_R(@128~ nm) = 1 ~m$ \\ $\rm \lambda_R(@176~ nm) = 8.5 ~m$\end{tabular}} & {\begin{tabular}[c]{@{}c@{}} $\rm \lambda_{abs}(N2@128 ~nm) = 20 ~m$\\
$\rm \lambda_{abs}(N2@176~nm) = 80 ~m$ \end{tabular}} \\\hline\hline
 Parameter & PD efficiency & $\rm R_{FC}$ & $\rm R_{Cryostat}$ & $\rm R_{Anode}$ \\ \hline
 Value & 2\% & 79\% & {\begin{tabular}[c]{@{}c@{}}$\rm 30\% @128 ~nm$ \\ $\rm 40\% @176 ~nm$ \end{tabular}} & {\begin{tabular}[c]{@{}c@{}}$\rm 6\% @128 ~nm$ \\ $\rm 12\% @176 ~nm$ \end{tabular}} \\ \hline
\end{tabular}
\end{table}

\begin{figure}[htbp]
\centering
\includegraphics[width=.7\textwidth]{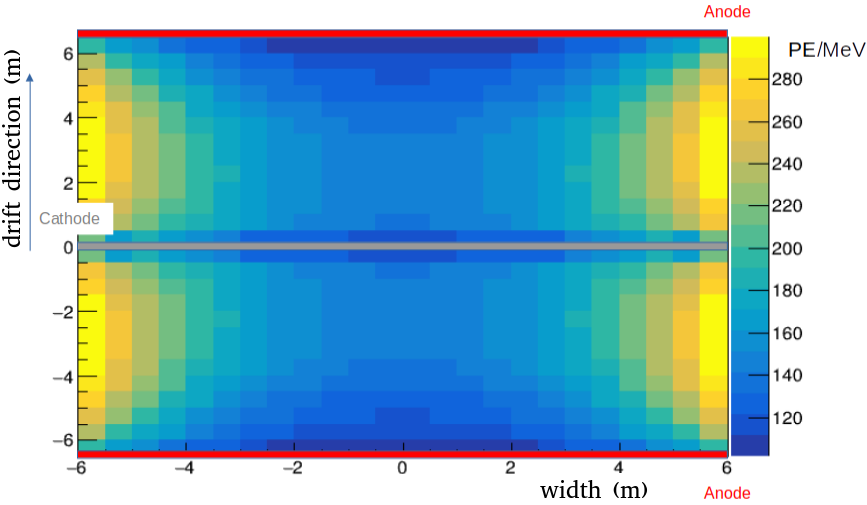}
\caption{Expected APEX light yield map at z=0 (with z the beam direction). A cross-section of the detector on the drift-width plane is shown.\label{fig:lymap}}
\end{figure}

Figure \ref{fig:lymap} shows the expected $LY$ map given in units of number of detected photoelectrons per deposited energy (PE/MeV) at the transversal plane $z=0$. The $LY$ map remains practically unchanged within the region $z=\pm \rm 20~m$. Outside such region $LY$ increases as the FC shorter sides are also instrumented with PD units. The fraction of light emitted at 176 nm due to Xe doping was taken as 53\% while the remaining LAr scintillation light is assumed to be reduced to 30\% of the total argon scintillation yield  \cite{Vdtdr, xedope}. Although the total number of photons produced is lower when Xe is present at minimal levels in the volume, the 176 nm light propagation characteristics allows light to reach PD units farther from the light source events resulting in higher expected $LY$ values and a more uniform map than with pure LAr. The detection of the backwards emitted pTP shifted light also has a large impact accounting for an increase of $\rm\sim60\%$ with respect to the immediate detected scintillation light. The average and minimum $LY$ are estimated as $ LY_{av} = \rm 180~PE/MeV$ and $LY_{min} = \rm 109~PE/MeV$ respectively in the central $z$ region. Comparing these numbers to the estimates obtained with similar simulation for the VD we observe an increase of the $LY_{av (min)}$ by a factor 4(6). 

The capability of 3D positioning of low energy interactions by the PDS was also studied. For that, simple coordinate estimators were employed:
\begin{equation}\label{posit}
    z_{rec} = z_{bar},~~~~~
    y_{rec} = L r (a+b|r|),~~~~~
    x_{rec} = c(y_{rec})+d(y_{rec})x_{bar}.
\end{equation}
The $z_{rec}$ coordinate was estimated as the PE weighted barycenter $z_{bar}$ obtained with the $z$ coordinates of the fired PD units in a given event. The $y_{rec}$ estimator provided the reconstructed horizontal coordinate, with the half width $L$ of the active LAr volume, $a$ and $b$ constants and $r$ given by:
\begin{equation}
r = \frac{N_{L+} - N_{L{-}}}{N_{L+}+N_{L{-}}},
\end{equation}
where $N_{L\pm}$ is the number of PEs on $x=\pm L$. The $x_{rec}$ vertical reconstructed coordinate was calculated via a straight line relation with respect to $x_{bar}$ being $c$ and $d$ linear functions of $y_{rec}$ themselves.

\begin{figure}[htbp]
\centering
\includegraphics[width=.49\textwidth]{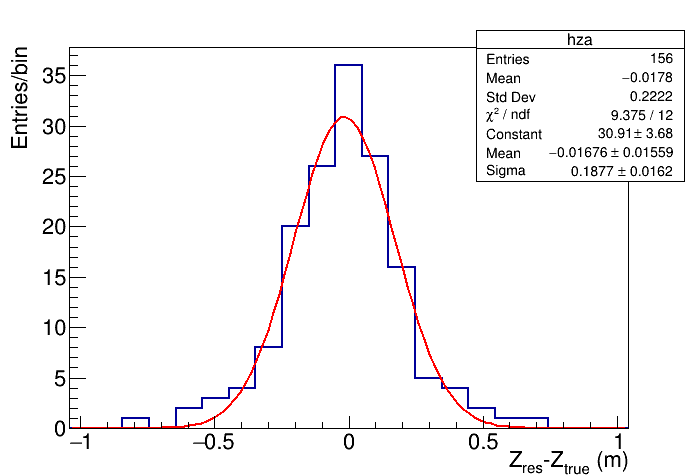}
\includegraphics[width=.49
\textwidth]{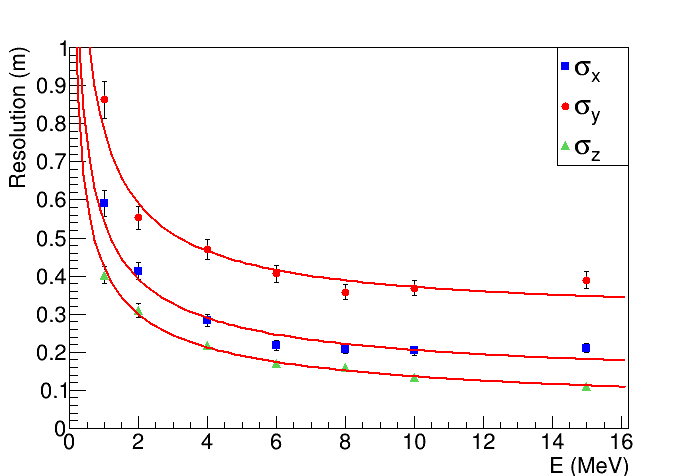}
\caption{$Z_{rec}$ residuals distribution of 4 MeV deposits (left). Position resolution per cartesian coordinate as function of the true values of the deposited energy. Red lines as the fits of the distributions: Gaussian on the left and $E^{-1/2}$ on the right.\label{fig:zres}}

\end{figure}

Figure \ref{fig:zres} shows the residuals distribution obtained for $z_{rec}$ (left) for 4 MeV energy deposits. The residuals distributions for all coordinates were found to be fairly uniform across the whole LAr volume showing no biases. The spatial resolution was also obtained as function of the deposited energy (right). Comparison with a similar simulation performed for VD \cite{vdsim} indicates that APEX spatial resolution is a factor $\sim0.5$ smaller.

\begin{figure}[htbp]
\centering
\includegraphics[width=.6\textwidth]{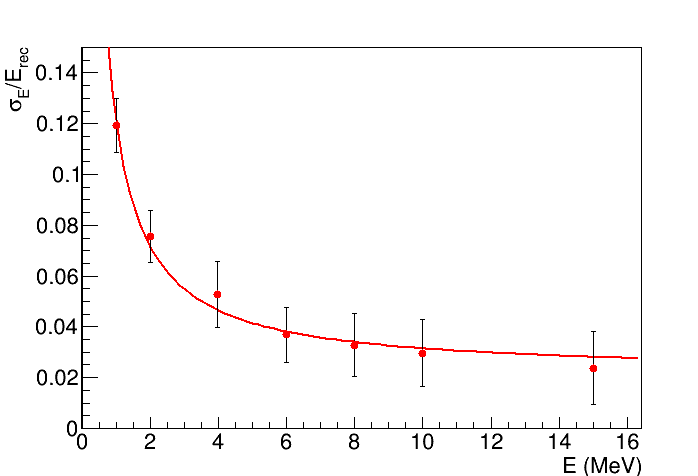}
\caption{Energy deposit relative resolution as function of its true value. The distribution is fitted (red line) by the quadratic sum of constant and stochastic ($\propto 1/\sqrt{E}$) terms.  \label{fig:eres}}
\end{figure}

Another performance quantity of interest is the energy deposit resolution. This resolution can be evaluated by comparing the true deposited energy with the reconstructed one. The reconstructed deposited energy is calculated as:
\begin{equation}
E_{rec} = \frac{\sum{PE}}{LY(x_{rec},y_{rec},z_{rec})},
\end{equation}
where $LY$ was calculated with the reconstructed energy deposit position and a map with segmentation size similar to the expected from calibration uncertainties. This is modeled with the coarse LY map shown in figure \ref{fig:lymap}, which is consistent with the one expected to be obtained experimentally (e.g. by pulsed neutron source calibration). Statistical fluctuations on individual PE values were introduced due to detection efficiency. Figure \ref{fig:eres} shows estimates for the relative energy deposit resolution as a function of the true energy value. The error bars represent the average variation of the obtained values across different positions in the cryostat. A comparison with VD estimates shows resolution on low energy deposits reduces by factor $\rm \sim0.4$ for APEX.

\section{Conclusions}

APEX is an optimized PD system based on DUNE's second far detector module, VD, R\&D achievements and new readout technology advancements. Preliminary Monte Carlo studies and comparisons to VD demonstrate it can deliver much superior performance in all detection aspects of interest, possibly impacting DUNE's physics capabilities considerably. The expected light yield is at least 4 times higher than the one expected for VD ($ LY_{av} = \rm 180~PE/MeV$ and $LY_{min} = \rm 109~PE/MeV$ vs  $ LY_{av} = \rm 39~PE/MeV$ and $LY_{min} = \rm 16~PE/MeV$ in the central region of the detector) thanks in part to the efficient collection of the light back-scattered of the pTP deposited on the acceptance window of photon-detector modules. The position resolution based on the PDS alone can reach values below one meter in any direction, getting as low as 10-20 cm  along the beam direction. Regarding the energy resolution, it is on the order of 3\% for tens of MeV deposited energy. Prototyping stages are ongoing and activities foreseen for the next years aiming for a full-sized APEX PD-instrumented field cage to be deployed in the VD ProtoDUNE cryostat at CERN.




\bibliographystyle{JHEP}
\bibliography{biblio.bib}

\end{document}